\documentclass[sigconf]{acmart}

\usepackage{algorithmic}
\usepackage{graphicx}
\usepackage{textcomp}
\usepackage{xcolor}
\usepackage{subfig}
\usepackage{multirow}

\AtBeginDocument{%
  \providecommand\BibTeX{{%
    \normalfont B\kern-0.5em{\scshape i\kern-0.25em b}\kern-0.8em\TeX}}}

\setcopyright{acmcopyright}
\copyrightyear{2022}
\acmYear{2022}
\acmDOI{10.1145/1122445.1122456}

\copyrightyear{2022}
\acmYear{2022}
\setcopyright{rightsretained}
\acmConference[WebSci '22]{14th ACM Web Science Conference 2022}{June 26--29, 2022}{Barcelona, Spain}
\acmBooktitle{14th ACM Web Science Conference 2022 (WebSci '22), June 26--29, 2022, Barcelona, Spain}\acmDOI{10.1145/3501247.3531542}
\acmISBN{978-1-4503-9191-7/22/06}

\begin{document}

\title{Online Coordination: Methods and Comparative Case Studies of Coordinated Groups across Four Events in the United States}


\author{Lynnette Hui Xian Ng, Kathleen M. Carley}
\email{huixiann,carley@andrew.cmu.edu}
\affiliation{%
 \institution{CASOS, Institute for Software Research, Carnegie Mellon University}
 \city{Pittsburgh}
 \state{Pennsylvania}
 \country{USA}}

\renewcommand{\shortauthors}{Ng and Carley}

\begin{abstract}
Coordinated groups of user accounts working together in online social media can be used to manipulate the online discourse and thus is an important area of study.
In this study, we work towards a general theory of coordination. 
There are many ways to coordinate groups online: semantic, social, referral and many more. Each represents a coordination dimension, where the more dimensions of coordination are present for one event, the stronger the coordination present.
We build on existing approaches that detect coordinated groups by identifying high levels of synchronized actions within a specified time window. A key concern with this approach is the selection of the time window. We propose a method that selects the optimal window size to accurately capture local coordination while avoiding the capture of coincidental synchronicity.
With this enhanced method of coordination detection, we perform a comparative study across four events: US Elections Primaries 2020, Reopen America 2020, Capitol Riots 2021 and COVID Vaccine Release 2021. 
Herein, we explore the following three dimensions of coordination for each event -- semantic, referral and social coordination -- and perform group and user analysis within and among the events. This allows us to expose different user coordination behavior patterns and identify narratives and user support themes, hence estimating the degree and theme of coordination. 
\end{abstract}

\ccsdesc[500]{Human-centered computing~Social network analysis}

\keywords{coordination detection, community clustering, narrative analysis, URL analysis, social network analysis
}

\maketitle

\section{Introduction}
Social media has facilitated the propagation of news information to the masses, but has at the same time been a hotbed for groups of accounts deliberately spreading disinformation or performing influence campaigns \cite{doi:10.1126/science.aao2998,Bessi_Ferrara_2016}. 
Coordinated groups of user accounts working together to manipulate the online discourse is an important area of study for social cybersecurity \cite{weber2021temporal}.  
Much of the behavior of coordinated user groups are innocuous at the individual level and require further scrutiny of network interactions among accounts over time \cite{Pacheco_Hui_Torres-Lugo_Truong_Flammini_Menczer_2021}. 
For example, a single user constantly tweeting a hashtag may appear harmless but a group of users tweeting a series of hashtags within a small timeframe of each other raises suspicions of coordinated activity through synchronization of behavior.

Studies have been performed on how coordinated groups artificially manipulate online information on elections.
\citet{giglietto2020takes} recognized that it takes a village to manipulate the media in their study of 2018-2019 Italian elections.
Through analyzing Facebook co-shares of political news stories, they identified hundreds of groups that coordinated to boost political and non-political identities.
Coordinating groups on social media also pose a threat to social fabric, especially when their campaigns spill over into the offline medium and result in protests and riots.
\citet{Steinert-Threlkeld2015} used hashtag coordination to capture the noticeable change between online coordination and offline protests in 16 countries affected by the 2011 Arab Spring protests.
Analysis of similar images shared in the 2019 Hong Kong protests results in a polarizing network, namely pro- and anti-protest users \cite{Pacheco_Hui_Torres-Lugo_Truong_Flammini_Menczer_2021,ng2021coordinating}, while analysis of similar texts in the 2021 Capitol Riots reveals user clusters supporting themes of disinformation narratives \cite{ng2021coordinating}.

Online coordination is a multi-dimension problem. 
Current coordinated activity detection techniques typically uncover anomalously high level of synchronized action within a time window. An action is a behavior a user can take on a social media platform such as retweets \cite{9381418}, @-mentions, \cite{DBLP:journals/corr/abs-2105-07454}, using similar texts \cite{ng2021coordinating}, posting common URLs \cite{Bao2018,cao2015organic}, or other behavioral-traces that links two users \cite{Pacheco_Hui_Torres-Lugo_Truong_Flammini_Menczer_2021}. 
A key concern with this approach is the selection of the time window in defining coordinating actions. 
We further the study in this field by evaluating the choice of an optimal window size with the definition of synchronized actions for measuring local coordination without capturing coincidental synchronicity. 
With this definition, we analyze four major events in the United States that took place from 2020 to 2021. 
We uncover coordinating communities through semantic, referral and social coordination. We perform text and URL analysis to systematically characterize the themes propagated and perform user analysis both within and across the events, providing a comprehensive analysis of coordinating clusters.



\paragraph{Contributions.} Our main contributions are:  
\begin{enumerate}
    \item A methodology for the discovery and analysis of coordinated groups through uncovering anomalously high levels of synchronized actions. This methodology builds on the Synchronized Action Framework \cite{DBLP:journals/corr/abs-2105-07454} by further defining optimal time windows. It can in principle be applied to any social media platform where data is available. This method is completely unsupervised, hence no labeled training data is required.
    \item Using Twitter data, we present four case studies by instantiating the methodology to detect different types of coordination: (i) semantic coordination based on hashtags, (ii) referral coordination based on URLs and (iii) social coordination based on @-mentions.
    The methodology also involves characterizing the themes within coordinated groups, which may be interpreted as the coordinated campaign, based on the narrative of the text written, the URL content and the screen name.
    \item Using the discovered coordinated groups, we perform user analysis within the groups and across the events, further understanding the types of accounts participating in coordinated activity and their behavior within the four communities. Understanding cross-type and cross-event coordination provides an estimation of the degree coordination. 
\end{enumerate}

\section{Related Work}
Existing methods in the detection of coordinated groups rely heavily on identification and exploitation of behavioral traces or action synchronized in time \cite{DBLP:journals/corr/abs-2105-07454,Pacheco_Hui_Torres-Lugo_Truong_Flammini_Menczer_2021}.
These actions range from simple to complex.
Simple actions like shared URLs, hashtags or retweets can be derived from the collected data itself.
There has been success with this method in uncovering networks using simple shared actions: uncovering foreign interference in narrative boosting with co-retweets during the 2016 US Presidential elections \cite{9381418}, discovering groups polarizing the 2019 Hong Kong protest \cite{Pacheco_Hui_Torres-Lugo_Truong_Flammini_Menczer_2021} through similar images, identifying a group of partisan accounts boosting a single narrative in the 2018 Australian election through synchronized retweets \cite{weber2021temporal}, and uncovered bot networks that spread conspiracies during the COVID pandemic using co-retweets \cite{graham2020like}.

Complex actions require additional processing steps prior to analyzing synchronized actions. Case studies using complex actions include: analysis of similar texts with text vector similarity methods uncover disinformation themes in the Capitol Riots \cite{ng2021coordinating}, and  analysis of similar images with image-RGB-vector similarity reveals a pro- and anti- Hong Kong Protest segregation \cite{Pacheco_Hui_Torres-Lugo_Truong_Flammini_Menczer_2021}. Yet others require more user observations, such as in the case of identifying clickstreams  \cite{adda89c35bdb4c708af9401c5fb9c821}. 
Even more complex action modeling accounts for the temporal coincidence of organic synchronization through temporal point processes and Gaussian Mixture Models. This method uncovered groups of coordination surrounding anti-vaccination and anti-masks conspiracies in the COVID-19 pandemic \cite{10.1145/3447548.3467391}. Some of these groups are linked to state-sponsored operations, which is a worrying aspect of information operations, potentially leading into information warfare between countries \cite{woolley2018computational}.

After obtaining coordination networks, the next step is finding characteristic behaviors from the network structure. Techniques to group the coordinating users include graph clustering \cite{DBLP:journals/corr/abs-2105-07454}, extraction of highly coordinated communities \cite{9381418} and aggregation of users based on coordination weights \cite{10.1145/3447548.3467391}. Due to the large size of the resultant network, some analysis only investigate the largest coordinating cluster \cite{Pacheco_Hui_Torres-Lugo_Truong_Flammini_Menczer_2021,DBLP:journals/corr/abs-2105-07454}. To interpret these clusters of users, past studies typically use frequency count of actions and manual observations. For example, counting the frequencies of unique hashtags among coordinated accounts in the COVID19 pandemic reveals that coordinated groups promote anti-mask, anti-vaccine and anti-science theories \cite{10.1145/3447548.3467391}.
From manual URL labelling, coordinated URL group analysis reveals that coordinated organized groups only share from a median of 10 URL domains \cite{cao2015organic}. 


While these approaches can work well, each is designed to consider only one of the many possible coordination dimensions over one event. We further the study of online coordination by performing detection and analysis across multiple events, investigating the user behavior patterns across different communities.

\section{Data and Methodology}
\begin{figure}[!tbp]
    \centering
   \includegraphics[height=1.0\linewidth]{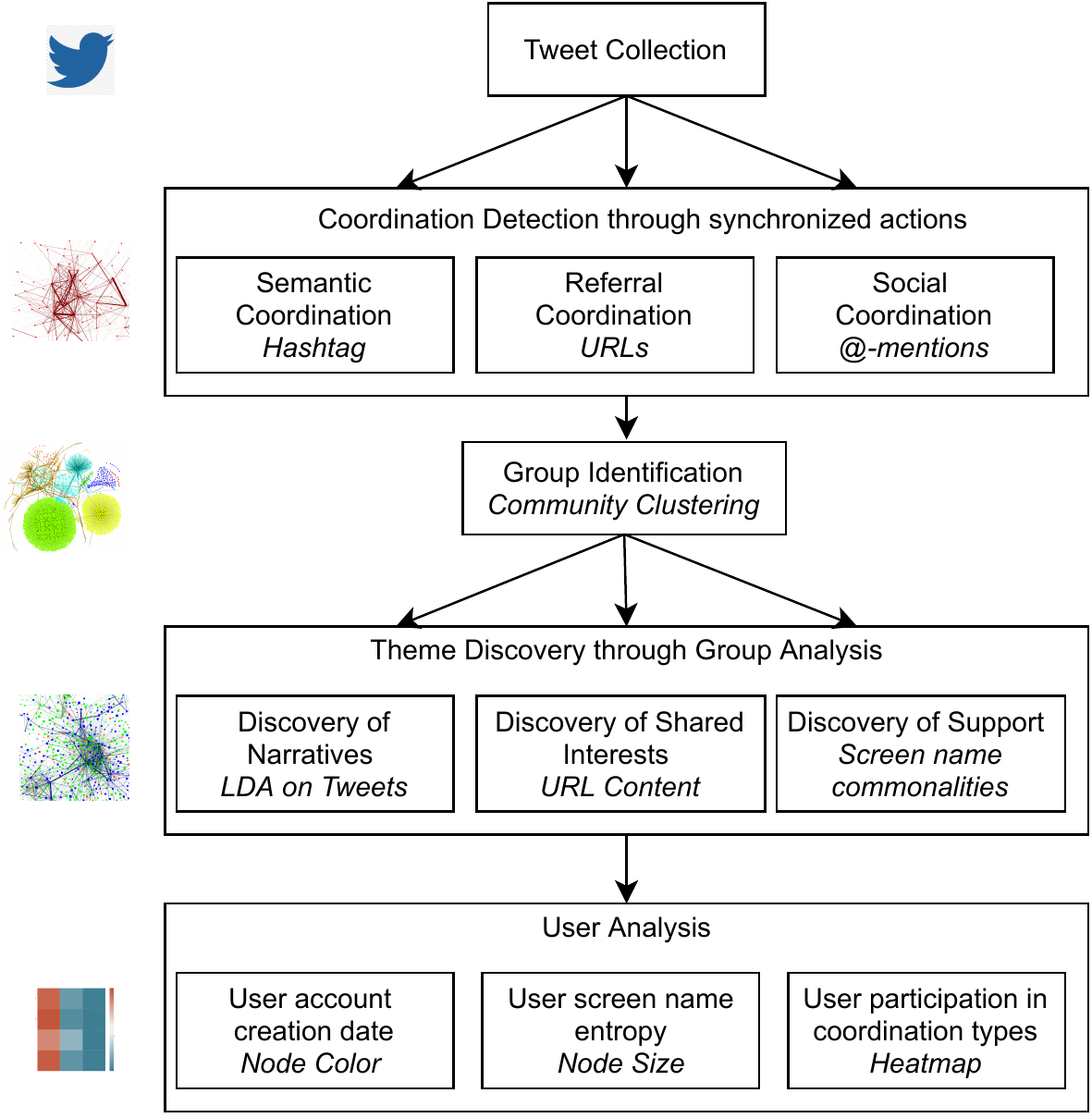}
  \caption{Methodology for the discovery and analysis of coordinated groups}
  \label{fig:pipeline} 
\end{figure}

In this section, we present the data collection procedure and the methodology used in the discovery and analysis of coordination in our study. A brief overview is presented in a flowchart in \ref{fig:pipeline}.
Our methodology harnesses network analysis as an effective representation of coordinated activity. Network community detection provides a natural way to detect clusters of coordinated groups. We supplemented the cluster detection with group and user analysis to characterize coordinated activity occurring in each event. 

\subsection{Tweet data Collection}
For this study, we collect data from Twitter across four major events that occurred in the United States. For each event, we first collected tweets using a defined collection hashtag with the Twitter V1 API. The events span from 2020 to 2021 and we ensured there are minimal overlap across the collection timeframes to reduce the number of overlapping tweets that may occur due to the use of more than one event hashtag. 

The four events that we analyzed in this study are: (1) US Elections Primaries 2020; (2) Reopen America 2020, in which protests termed ``Operation Gridlock" were launched across the United States against the government lockdown response to the coronavirus pandemic; (3) Capitol Riots 2021, in which a mob of supporters for then-President Donald Trump attacked the US Capitol building; (4) COVID Vaccine release 2021, in which a vaccine served as a light of hope for the global coronavirus (COVID) pandemic.
We filtered for original tweets (i.e. not retweet, quote tweet or replies) that were in the English language and were geomarked in the United States. 
Although filtering for original tweets greatly reduced the dataset size, original tweets represent tweets that originated from the user that was deliberately written. 
The summary of the collected data statistics is reflected in Table \ref{tab:dataset}.

\begin{table}[!hpt]
\centering
\begin{tabular}{|p{3cm}|p{1.8cm}|p{1.3cm}|p{1.3cm}|}
\hline
\textbf{Event} \newline \textbf{(Collection Hashtag)} & \textbf{Timeframe} & \textbf{Num Unique Users} & \textbf{Num Original Tweets} \\ \hline 
US Elections Primaries 2020 \newline \#uselectionsprimaries & 1 Feb 2020 - \newline 13 Feb 2020 & 628,705 & 2,223,056\\ \hline 
Reopen America 2020 \newline \#operationgridlock & 1 April 2020 - \newline 22 June 2020 & 88,768 & 200,631\\ \hline
Capitol Riots 2021 \newline \#stopthesteal & 1 Jan 2021 - \newline 7 Jan 2021 & 412,788& 763,850\\ \hline 
COVID Vaccine Release 2021 \newline \#covid \#vaccine & 9 Jan 2021 - \newline 9 May 2021 & 668,309 & 1,279,919\\ \hline 
\end{tabular}
\caption{Details of the datasets of the four events analyzed}
\label{tab:dataset}
\end{table}

\subsection{Coordination Detection}
We detect users that perform coordination with the Synchronized Action Framework\footnote{ \url{https://github.com/CASOS-IDeaS-CMU/coordination-analysis}} \cite{DBLP:journals/corr/abs-2105-07454} along three axes: (1) semantic coordination inferred through common hashtags, (2) referral coordination inferred through common URLs, and (3) social coordination inferred through common @-mentions tagging of other users.
We construct our coordination graph networks by detecting anomalously high levels of synchronized actions within a short time window \cite{DBLP:journals/corr/abs-2105-07454}. 
In our coordination graph networks, the nodes are represented by users. A link between two users represent the presence of a synchronized action within the time window, and the weight of the link represent strength of coordination, calculated by the number of times the action is performed within a time window. 

We construct three coordination networks for each event to investigate the three types of coordination. We thus define the three actions as: hashtags, URLs and @-mentions. Within a tweet, hashtags are identified by the \texttt{entities.hashtags} field, URLs are identified by the \texttt{entities.url.expanded\_url} field and @-mentions are identified by the \texttt{entities.user\_mentions.screen\_name} field. No preprocessing of the tweet texts were required as our methodology for coordination detection relied on tweet artifacts (hashtags, URLs and @-mentions). As a result, these artifacts were left intact in the tweets and used in the coordination detection algorithm.

\begin{figure}[!tbp]
  \centering
  \includegraphics[width=1.1\linewidth]{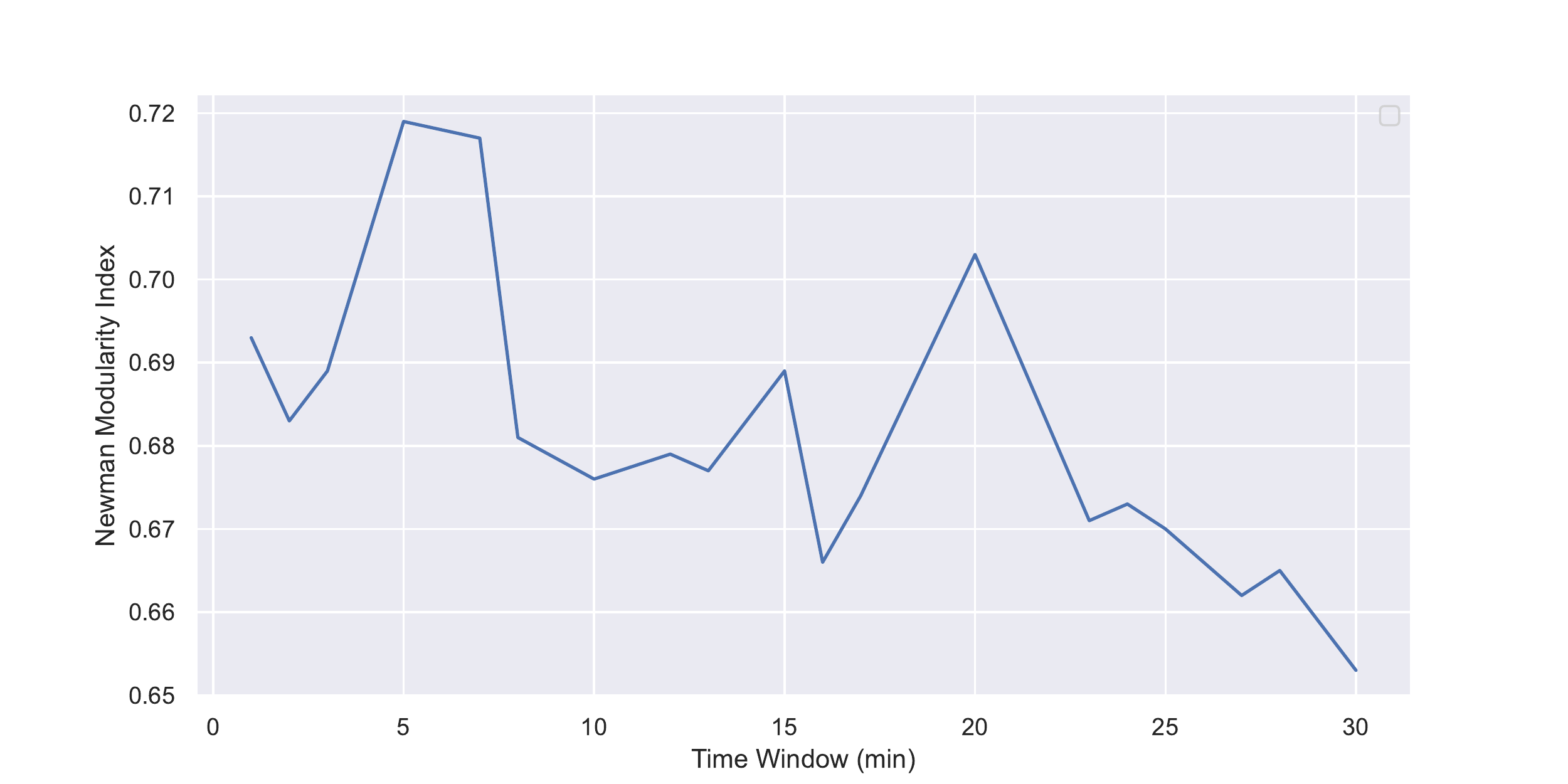}
  \caption{Average Newman Modularity of 1.3mil tweets in coordination network against Sliding Time Windows}
  \label{fig:newman} 
\end{figure}

In this step, the selection of the sliding time windows is crucial. A time window that is too short results in very little coordinating clusters, while a long time window captures coincidental synchronicity.
We determined the sliding window size through a preliminary experiment. Using half of the tweets in the US Election primaries, we formed the three coordination networks (semantic, referral, social) of different sliding window sizes ranging from 1 minute to 30 minutes.We then measured the average Newman Modularity Index.
The Newman Modularity Index is a measure of how well the network can be clustered \cite{newman2006modularity}. A high index is desirable as the network can be partitioned neatly into groups with high in-group links and low out-group links, while a low index results in a fragmented network.
Figure \ref{fig:newman} shows a peak of the Newman Modularity Index at the 5-min window before steady decrease of the modularity index, thus we use the 5-min sliding window for our coordination network formation.

Finally, we filter the network graph to retain users that coordinate with the link strength of \texttt{ceiling(mean + stdev)} of each graph. The removal of links below this value ensures that the threshold varies from each graph, taking into account different characteristics of each dataset. Doing so also reduces noise and thus retains only users that are synchronously performing the same action a high number of times.

\subsection{Group Identification}
After generating the initial coordination networks, we perform a thresholding on the link-weight of the networks, removing links that are below (mean + standard deviation) of the network's links. This thresholding method aids in reducing noise and sieving out the core coordinating users of the network.
We then use the Louvain clustering community detection algorithm on each network to identify groups of users. This clustering algorithm optimizes the node modularity which evaluates how densely connected the nodes are compared to a random network \cite{Traag2019}. 

\subsection{Theme Discovery through Group Analysis}
For each group in each network, we perform theme discovery in order to better understand the group. We describe the approaches taken to perform theme discovery corresponding to each coordination type.
\paragraph{Discovery of narratives from semantic coordination.}
For each Louvain group in a semantic coordination network, we obtain the tweets involved in the network formation. We pre-processed the tweet texts to remove punctuation, stop words and tweet artifacts (i.e. hashtags, @-mentions, URLs as these artifacts were already used for coordination detection), then perform a Latent Dirichlet Allocation using the sklearn library \footnote{https://scikit-learn.org} to derive one narrative cluster and manually interpret the terms to identify the narrative of the cluster, primarily using phrases that occur most frequently in the texts as the cluster's main narratives.

\paragraph{Discovery of shared interests from referral coordination.}
For each Louvain group in a referral coordination network, we obtain the URLs involved in the network formation. 
For each URL, we removed the base domains and retain the content. 
We pre-process the content as texts by removing punctuation and special characters and cluster the URL contents as term counts. For example, for the URL \textit{https://uscouriertoday.com/twitter-locks-president-trump-for-the-first-time/}, we retain the content as ``twitter locks president trump for the first time".
We manually interpret the terms to identify the shared interests within the cluster, primarily using phrases that occur most frequently in the texts from the cluster as the cluster's main interest.

\paragraph{Discovery of support from social coordination.}
For each Louvain group in a social coordination network, we obtain the screen names that were mentioned between pairs of users. 
We perform a manual inspection to identify common substrings among the screen names. Using this list of substrings, we find the frequency of the substrings appearing in the set of screen names that were being mentioned and manually interpret the high frequency substrings to identify themes users in the cluster support. For example, for users with the screen names ``ilovebubbletea" and ``bubbleteaislove", we derive the common substring to be ``bubble tea" and infer the users' support for bubble tea from the public proclamation via their screen name.

\subsection{User Analysis}
We performed (1) analysis of user account creation dates and (2) account screen name entropy to further analyse users in a cluster. 
\paragraph{Analysis of user accounts creation dates.} We visited each dataset six months later and collected user profiles using the Twitter API, in particular the dates the profiles were created. We also noted profiles which Twitter returned an error message that the profile was suspended.
We then annotate each user profile based on its creation date: suspended users, users created less than 3 months of the event, users created between 3-6 months of the event and users created more than 6 months of the event.
We overlay this information on to the coordination network graphs and further analyze the users that propagate each theme in the graph. 

\paragraph{Account screen name entropy.} We analyzed each account screen name in terms of its string entropy, which represents how probable the characters occur beside each other in a string. 
To quantify screen names used in Twitter, we collected 3,775,972 screen names of Twitter users who posted content from 2020 to 2021. Using this screen name corpus, we constructed a dictionary representing the probability of all unique characters (alphabets, numbers, punctuation etc) occurring in the corpus.
We elected to construct our own character probability distribution instead of using available distributions from the English language \cite{english} as often used in cryptographic texts because the formation of screen names on a social media platform can be different from formation of words in a language. 

We then calculate the entropy $H(X)$ of each screen name. For each $x_i$ of the $n$ characters in a screen name $X$, we obtain the probability $P(x_i)$ from our constructed distribution and calculate the entropy of the screen name in Equation \ref{eq:entropyequation}:
\begin{equation}
    H(X) = -\sum^n_{i=1}{P(x_i)}log_2P(x_i)
    \label{eq:entropyequation}
\end{equation}

We combine the information about screen name entropy with user account creation dates to characterize trends between the two variables.

\paragraph{User participation in coordination types.} 
We analyze the user participation in coordination types by first looking at the number of coordination that each user participates in. We analyze the volume and types of accounts that participate in more than one coordination type. 
We then zoom in to users that participate in two types of coordination and identify the proportions in the types of multi-way coordination the users coordinate in. In particular, we investigate combinations of two-way coordination because the proportion of users that take part in three-way coordination is extremely low.  
Lastly, we investigate unique users that participate in one of the three types of coordination across the events. We construct a four-way venn diagram to represent the information where users appear in more than one event. 

\section{Results}
\begin{figure*}[!tbp]
  \centering
  \includegraphics[width=1.0\linewidth]{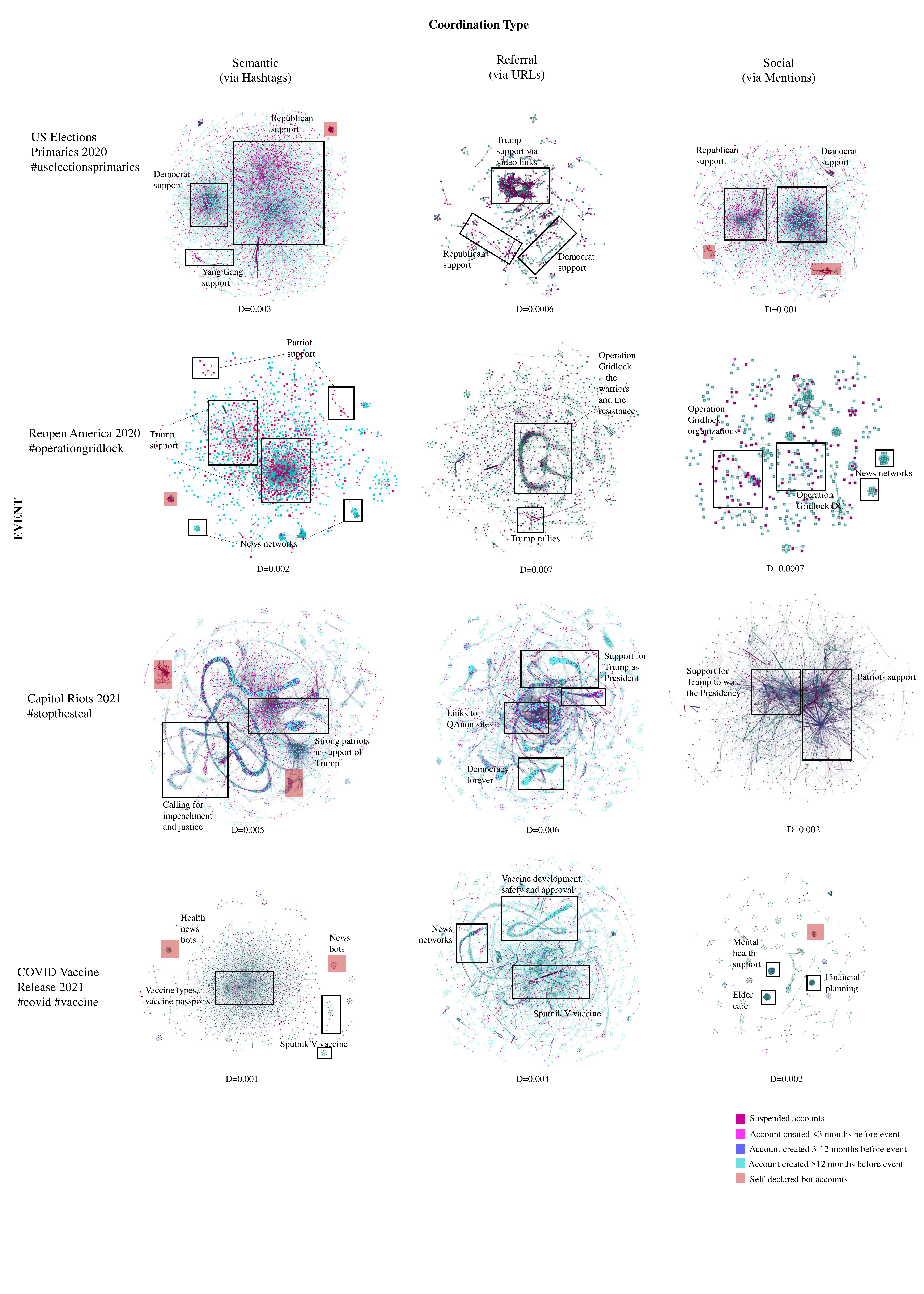}
  \caption{Coordination Analysis Results with network density D. Nodes are Twitter users and link width represent the strength of coordination between two users. Read boxes highlight clusters of users that have ``bot" in the user's screen name. }
  \label{fig:results} 
\end{figure*}

Through our coordination detection algorithm, we consistently discover coordinating groups among the four events studied. 
We first present the statistics of the number of nodes and links that were retained by the coordination detection algorithm as synchronized users in Table \ref{tab:filtered}.
We then visually show the final filtered network graphs in Figure \ref{fig:results}, presenting the coordination network diagrams with annotations of the discovered themes and the nodes are colored to reflect user account creation dates.

\begin{table*}[!hpt]
\centering
\begin{tabular}{|l|l|l|l|l|}
\hline
\textbf{Event} & \textbf{Stage} & \textbf{Semantic Coordination} & \textbf{Referral Coordination} & \textbf{Social Coordination} \\ \hline
\multirow{2}{*}{US Elections Primaries 2020} & Coordinated & Nodes:430k, Links: 1.92mil & Nodes: 19k, Links: 227k & Nodes: 127k, Links: 720k\\
& Filtered & Nodes: 5143, Links: 97k & Nodes: 2087, Links: 2962& Nodes: 2987, Links: 12k\\  \hline 
\multirow{2}{*}{Reopen America 2020} & Coordinated & Nodes: 59k, Links: 591k & Nodes: 30k, Links: 92k & Nodes: 50k, Links: 375k \\ 
& Filtered & Nodes: 1795, Links: 5525 & Nodes: 453, Links: 1187& Nodes: 3175, Links: 7717\\ \hline 
\multirow{2}{*}{Capitol Riots 2021} & Coordinated & Nodes: 98k, Links: 1mil & Nodes: 83k, Links: 1.53mil & Nodes: 54k, Links: 1.26mil \\ 
& Filtered & Nodes: 4969, Links: 124k & Nodes: 1323, Links: 4419& Nodes: 4511, Links: 117,01\\ \hline 
\multirow{2}{*}{COVID Vaccine Release 2021} & Coordinated & Nodes: 161k, Links: 1.83mil & Nodes: 83k, Links: 1.67mil & Nodes: 62k, Links: 1.01mil \\ 
& Filtered & Nodes: 1718, Links: 29k & Nodes: 7742, Links: 14k & Nodes: 3401, Links: 154k\\ \hline 
\end{tabular}
\caption{Statistics of stages in the coordination detection algorithm.}
\label{tab:filtered}
\end{table*}

\paragraph{Semantic Coordination}
In each of the four events, we identified key main narrative threads. There were two key narratives in the US Elections Primaries 2020: the Democratic support; the Republican support, especially on \#maga (Make America Great Again). There was a side narrative on Yang Gang, a devoted Democratic supporter who campaigned early in the US Presidential Elections race, characterized by 49 users, of which 50\% were created 3 months prior to the event. 39.7\% of users that participated in the Republican discourse were suspended by the platform six months later.

Reopen America 2020 network mainly included narratives that support then-President Trump. 36\% of these accounts were suspended six months later. 
Semantic coordination also sieved out groups of news networks from the same country who were likely to be Tweeting their regional news at the same time. 

The two largest clusters of narratives in the Capitol Riots 2021 called for justice and impeachment of then President Trump, while another cluster were ``patriots'' that supported Trump. 52 users formed a self-declared bot network and all users were suspended.

COVID vaccine discourse was dominated by the debate between the Pfzier, Moderna and Astrazeneca vaccines developed in response to the pandemic, and vaccine passports. Another small group of 25 users discussed the spread of Russian disinformation but were mainly discussing within themselves.
The methodology revealed groups of health and news bots and a group of 10 users whose names started with four `o' characters. Most accounts were created before 2020 (72.1\%) and only 13.8\% were eventually suspended.

\paragraph{Referral Coordination}
The key themes in the US Elections is promotion for Republican candidate Donald Trump using links to YouTube and Periscope videos, in particular the successful Trump-India trip. 60.9\% of these user accounts were later suspended.

ReOpen America 2020 had sites promoting the rallies of Republican candidate Trump and themes on the being the warriors and the resistance during the pandemic lockdown.

The Capitol Riots had three key themes of sites: democracy forever, links to QAnon sites and support for Donald Trump as president. 21.2\% of users were subsequently suspended. However, in the cluster that promoted democracy forever, 91\% of the users were created more than a year before the riot in 2021.

The COVID vaccine release event revealed links to three main topics of discussion: general news on the pandemic, news on vaccine development, safety and approval, and news on the Russian Sputnik V vaccine. 84\% of these accounts were created at least a year prior (before 2020). On the Sputnik vaccine, the two users with the highest coordination value and their neighbors were created in 2021, less than three months before the data collection.

\paragraph{Social Coordination}
US Election Primaries revealed users that overtly support Trump and Biden, the two key candidates for the Presidential race. The Democrat support was three times more than the Republican support. These users put the terms in their screen names like ``DefendDemocraC'' and ``TrumpIsMagic''. Interestingly, only 2.1\% of Republican supporters were later suspended while 33.6\% of Democrat supporters were suspended. 
Reopen America showed clusters of mentions of users that reflect some form of organizing the \#operationgridlock protests in their screen name. In these clusters, although no accounts were suspended, 42\% of the accounts were created within three months to the event.
There were also clusters of news networks that were reporting the protests. 

The Capitol Riots revealed a tightly meshed social support network, with users declaring themselves as patriots and purporting that Trump will win the Presidential race. 

The COVID vaccine revealed sparse isolated clusters of social coordination, mainly related to finance, elder care and mental health. 77.6\% accounts were created at least a year before and only 12.3\% accounts were suspended.

\paragraph{User analysis.}

\begin{figure*}
    \subfloat[\textbf{Screen name entropy by account type}]{\label{fig:entropy}\includegraphics[width=1.0\textwidth]{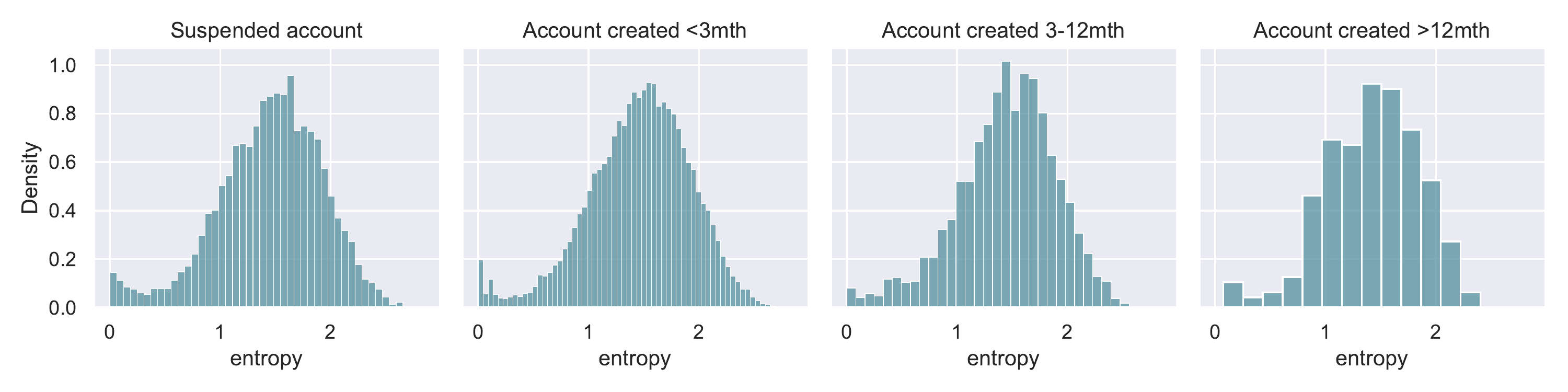}} \\
    \subfloat[\textbf{Strength of coordination by account type}]{\label{fig:strength}\includegraphics[width=1.0\textwidth]{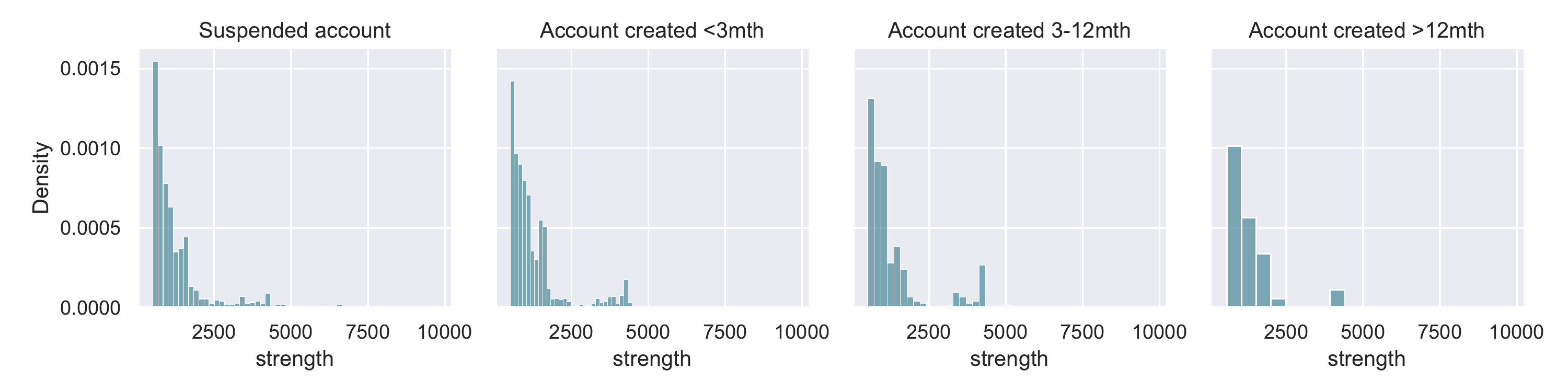}}
    \caption{Screen name entropy and strength of coordination by account type}
    \label{fig:entropyandstrength}
\end{figure*}

Clusters where users have high entropy and were suspended after six months typically relate to harmful themes, like electoral voter fraud support and the Russian vaccine conspiracy theory. In Figure \ref{fig:results}, we highlight clusters have that have ``bot'' in the user's screen name using red boxes, suggesting the overt representation of a bot network. 
We see that most users are created less than three months prior to the event, and there is a large number of users that are suspended in the four datasets we collected. 
We observe that the Twitter platform suspends accounts that propagate harmful narratives (e.g. patriot support) within six months of the Tweet collection, indicating that the platform successfully detects these users.
We also analyzed the screen name entropy and strength of coordination of each user (Figure \ref{fig:entropyandstrength}). In this, we see that users that are created less than three months prior to the event have the largest spread of screen name entropy at 1.42$\pm$0.43 bits, and the largest spread of strength. While screen name entropy is Normally distributed across all user types, coordination strength is generally right-skewed.  

\begin{figure*}[!tbp]
  \centering
  \subfloat[\textbf{By number of coordination types per event}]{\includegraphics[width=0.35\textwidth]{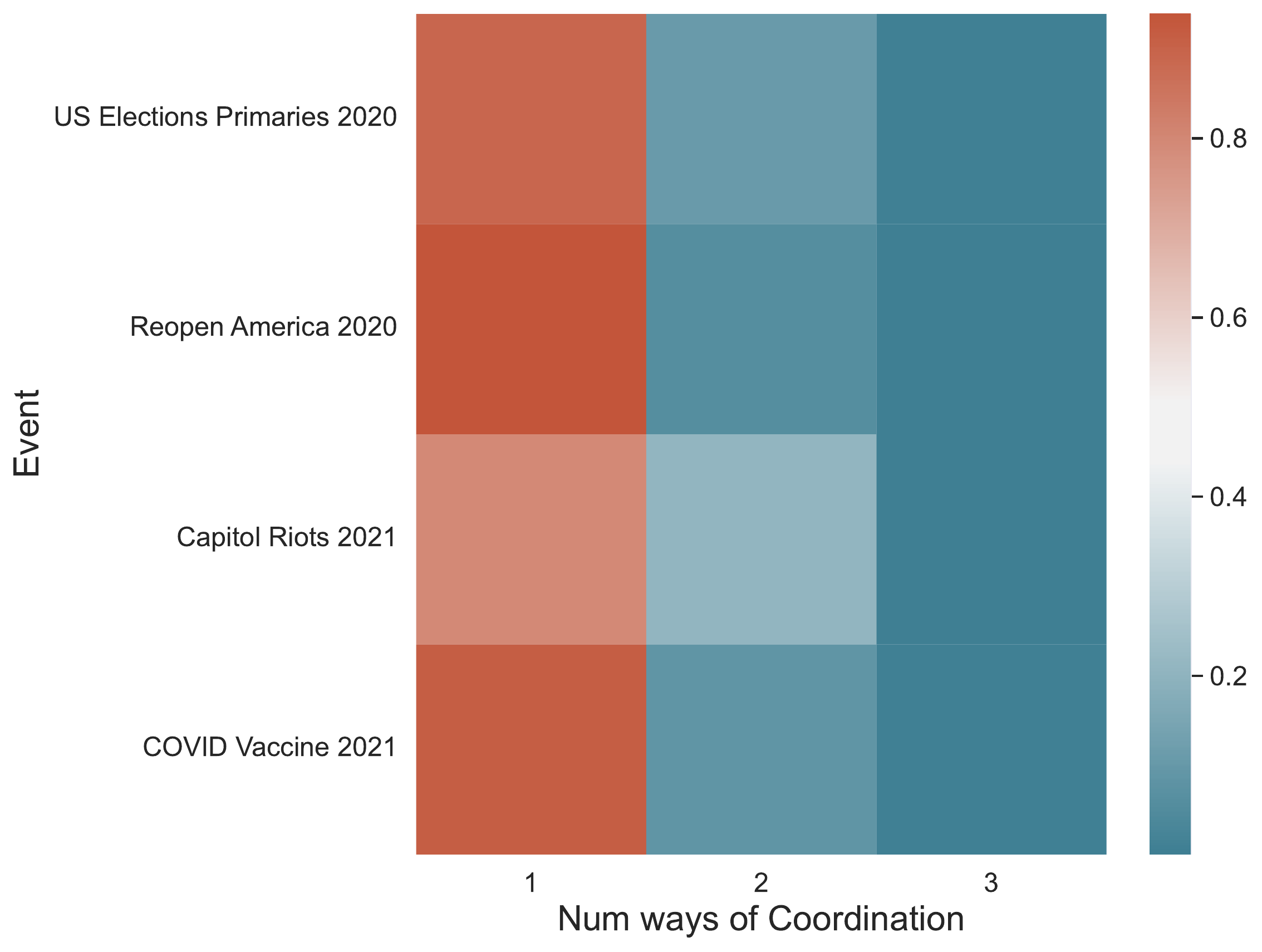}\label{fig:numwayscoordination}}
  \centering
  \subfloat[\textbf{By two types of coordination per event}]{\includegraphics[width=0.35\textwidth]{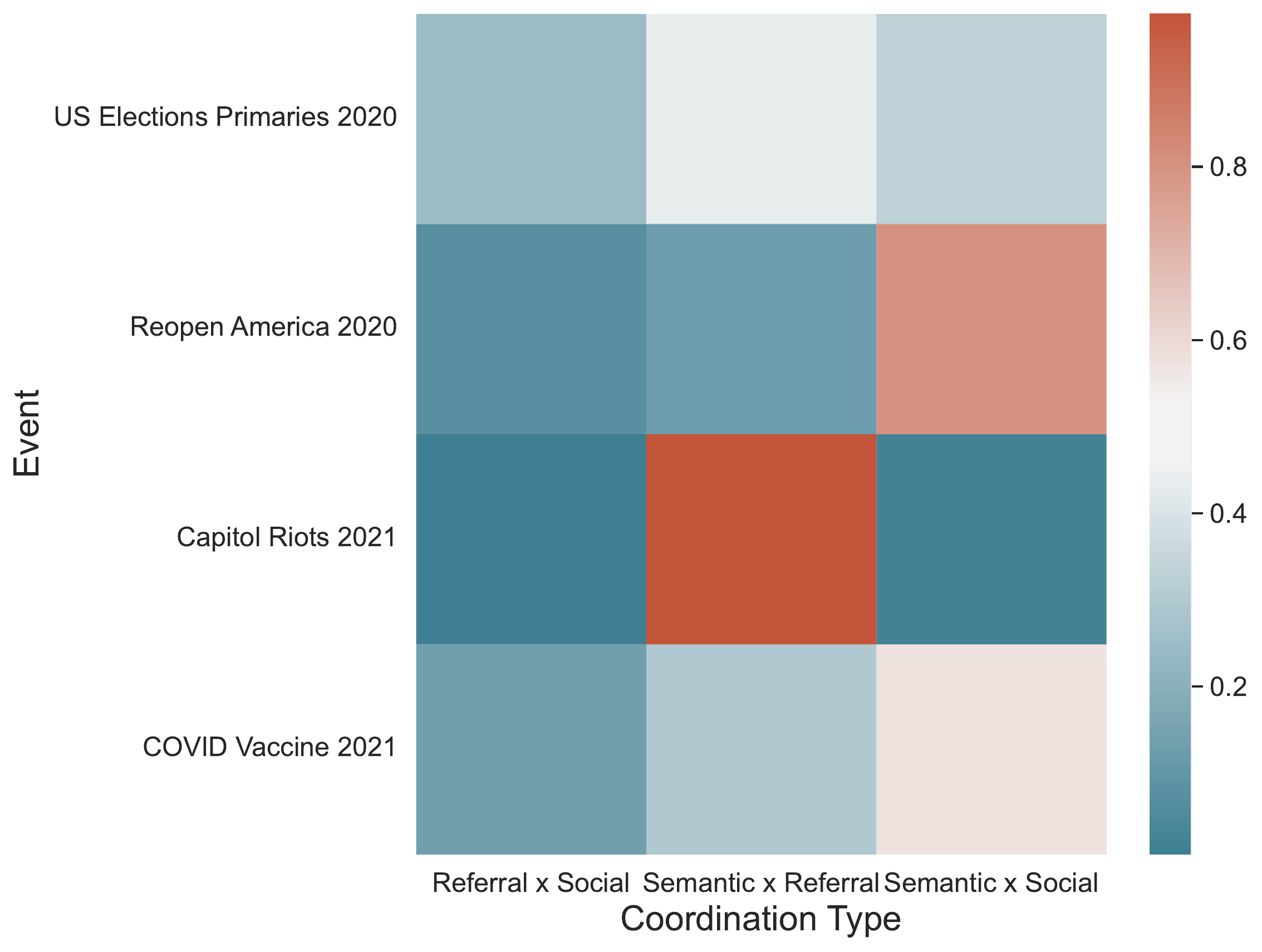}\label{fig:crossover_tabulation}}
  \centering
  \subfloat[\textbf{By event}]{\includegraphics[width=0.3\textwidth]{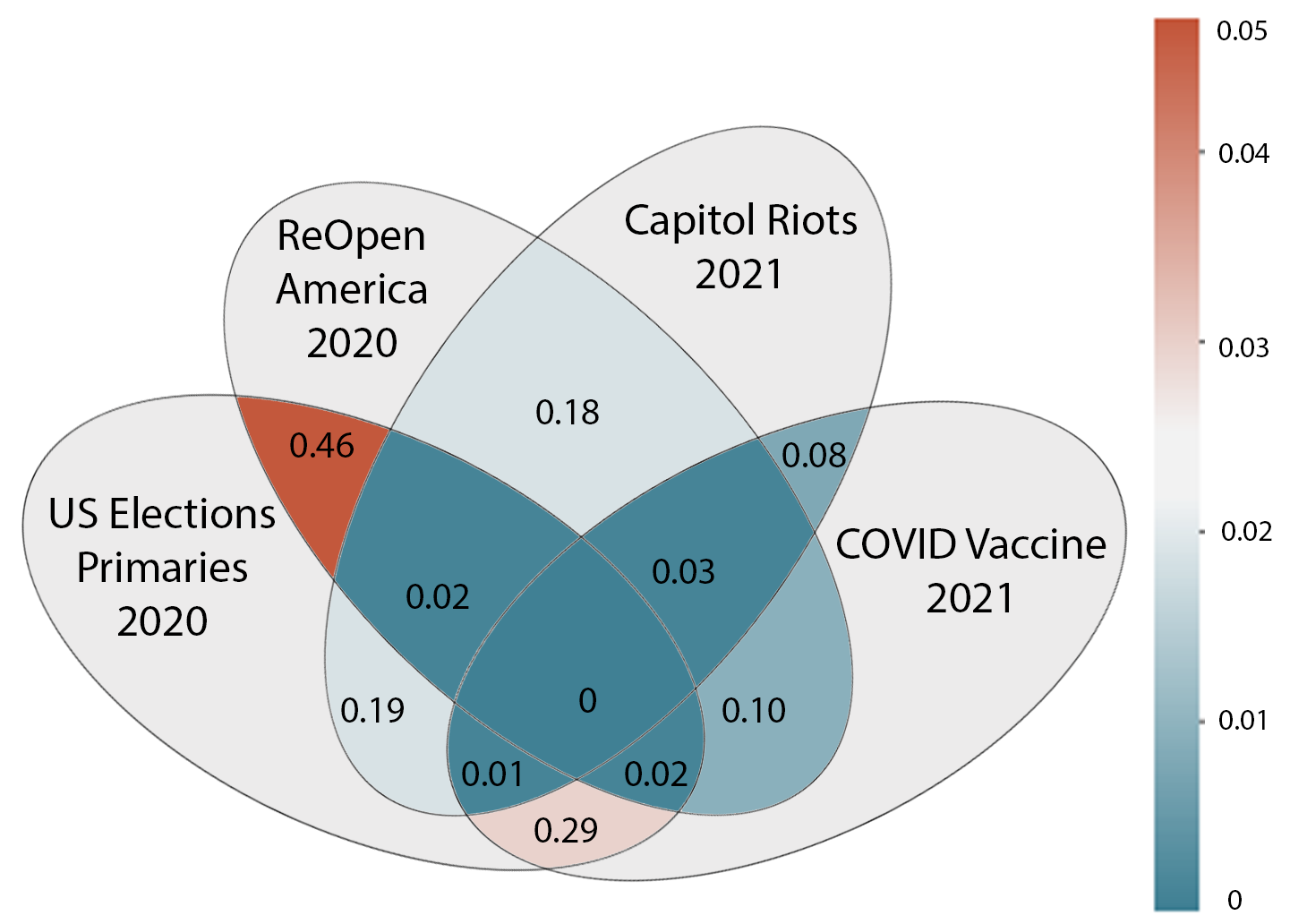}\label{fig:venndiagram}}
\label{fig:crossover}
\caption{User Analysis of the proportion of the users participating in multiple types of coordination.}
\end{figure*}

Our case studies not only explores single behaviors in isolated contexts. 
It also evaluates if certain types of coordination are related and the users types that participates in them. 
A handful of users engage in multiple types of coordination within the same event (Figure \ref{fig:numwayscoordination}). This percentage, while small ($\leq0.8\%$), is still noteworthy because these numbers are huge at a large scale. For example, the proportion of users engaging in all three types of coordination in the COVID Vaccine event is 0.1\%, which translates to 1200 users. This is a scale that can potentially do harm, given that a recent study on open-source datasets of bots discussing COVID19 range from 48 to 1427 users \cite{info:doi/10.2196/26933}.

We next turn our attention to users that engage in two coordination types. 
The proportion of users participation in two coordination types are visualized in a cross-tabulation table in Figure \ref{fig:crossover_tabulation}.
The proportion of users engaging in two coordination types is $0.055\pm0.002$, and the proportion engaging in three coordination types is $0.0035\pm0.0021$.
The screen name entropy of users that engage in 1-type of coordination is significantly less than users engaging in 2-types (p=0.004) and 3-types (p=0.004) of coordination. For users that coordinate 2-ways, the users that were later suspended had higher strength of coordination (p=0.002).



We further observe that there are a handful of users that engage in coordination in multiple events, presented in the Venn diagram in \ref{fig:venndiagram}. The number of users that participate in more than one event is very small ($<1\%$). The largest overlap of the number of users is between the US Election Primaries and ReOpen America events. 
In users that participate in two events, 19.9\% of the users were suspended and 77\% were created for more than 12 months. We observe a similar trend in users that participate in 3 events: 12.5\% of the users were suspended and 87.5\% were created for more than 12 months. No users participated in all four events. 



\section{Discussion}
In this work, we present a pipeline of coordination detection and analysis among Twitter users in four events through synchronized actions. We discover themes around semantic, referral and social coordination appearing in the events.
The four case studies presented in this work are merely illustrations of how the methodology can be implemented to discover and analyze coordinated groups. In principle, the approach can be applied to other social media platforms besides Twitter through the content users share.

Themes form around users with high screen name entropy. While screen names are generated by Twitter during account creation, users can change their it themselves. We thus infer that screen names with high string entropy represent users that do not bother about changing screen names, in which case they can be part of a group of users generating many new accounts at once and thus are part of coordinated activity. 

Coordination activity analysis can also be used to discover bot networks. In our study, we looked mainly at self-identified bots, which are accounts that have the term ``bot'' in their screen name. Many of these accounts are revealed to form coordination clusters on their own, suggesting they work together to boost each other's messages. These may be helpful to bot discovery tools. However, in the subset of users that coordinate across events , only 4\% are news accounts and 1\% are self-declared bots. While a more extensive bot and news agency analysis should be run, this preliminary result points that users that coordinate in multiple events are likely to be organic users.  

\paragraph{Multiple-coordination and cross-event user analysis.}
We observed that users participate in both semantic and social coordination is in events where discourse is mainly isolated to the virtual space, like the COVID Vaccine 2021 discussion. In this event, we postulate the users @-mention other users to change their stance towards the event, or the COVID vaccine in this case.
Semantic and referral coordination is observed to be related in events that have an offline presence, like the Reopen America 2020 and Capitol Riots 2021 events. 46\% of the URLs referred to in the tweets in these events refer to videos or images of the real-world protests.
The least proportion of users participate in referral and social coordination as compared to the other pairs of coordination type. 

\begin{figure*}[!tbp]
  \centering
  \includegraphics[width=1.0\linewidth]{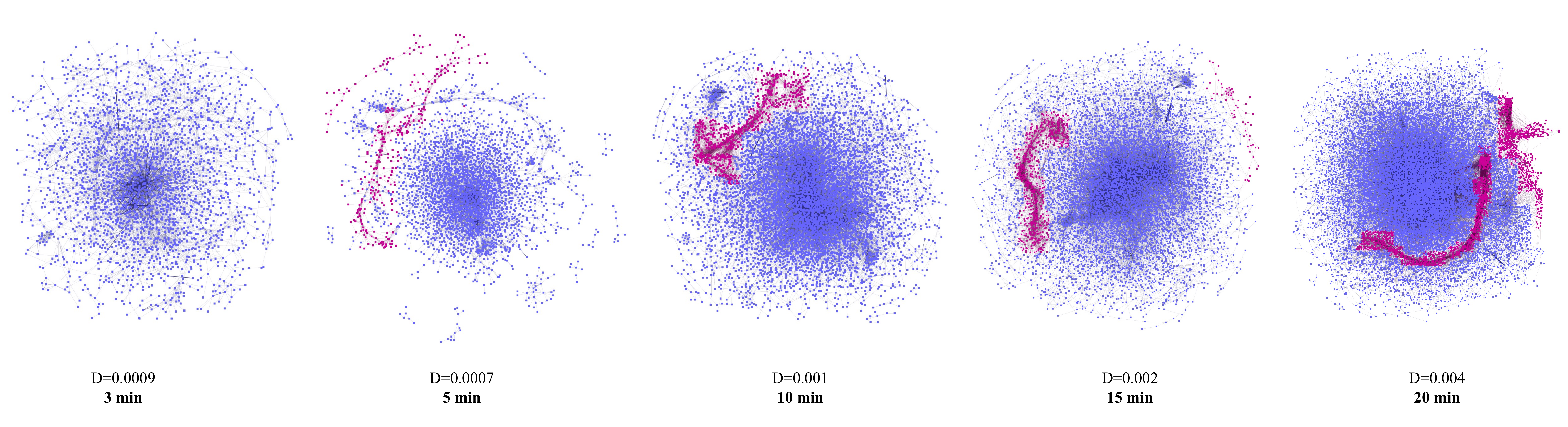}
  \caption{Semantic coordination graphs of across different time windows. Snaking network chains colored in red appear with denser networks which forms in larger time windows.}
  \label{fig:temporal_graph} 
\end{figure*}

In our cross-event user analysis, for accounts that perform coordination detection across events, a smaller number of accounts are suspended than the number of accounts that are still living on.
This also highlights that users that participate in more than one event may be organic rather than inorganic users. Past work have presented huge evidence of bot conversation and coordination on social media \cite{stefanidis2018bots,10.1007/978-3-319-93372-6_23}, but have also observed a shift from automated interference to self-organized social media distortions in 2020-2021 events of the US Elections, COVID pandemic, lock down and mask-wearing policies \cite{chang2021social}. This is consistent with our observations in users that coordinate across multiple events, suggesting that the social media space may increasingly be dominated by organic narrative and network manipulation. 
Additionally, observations about the screen name entropy may point to features for future investigation in prediction of user coordination activity.


\paragraph{Temporal Effect of Synchronized Actions.}
Our results show a phenomenon of snaking network lines across some coordination networks. These snaking network lines represent cascade chains of users coordinating with each other, e.g. A coordinates with B and B coordinates with C. We posit that these chains generally occur in dense network graphs with blurred divisive lines ($D>0.03$) For example, the chains are absent in the US Elections, where there is a strong partisan divide, but are present in the Capitol Riots where there are no clear stances. These snaking lines have also been observed in analysis of malware communities with dense networks ($D=0.022\pm0.015$) \cite{cruickshank2020analysis}. An empirical visual investigation with different time windows shows that chains form in larger time windows, because larger time window provides a larger group of users to use the same action within that window. 
Figure \ref{fig:temporal_graph} shows the semantic coordination network using half of US Election Primaries data revealing clear snaking network chains in denser networks which occurs in larger time windows. 
Closer inspection of the highlighted chain in pink reveals the same groups of users embedded within in growing chain: users discovered in the 5-min chain are found in the 10-min chain and so forth.
This is consistent with a previous study on temporal nuances in coordination networks \cite{weber2021temporal}, opening directions for future work to develop a systematic methods of identifying optimal temporal windows. These temporal windows might also be a function of the volume of the tweets at each time slice: a more active discourse might require a shorter time window otherwise the resultant network can be too dense for analysis. 
However, chains are almost always absent in the social coordination network, in which users tend to @-mention others in clear clusters, suggesting clear user support group distinctions.

\paragraph{Limitations.}
Twitter API limitations nuance the generalizability of our work. The API returns only 1\% of the Tweets, so there may be more coordinating actors and other discourse topics which were not captured in the dataset. 
This might have reduced the number of actors and groups found in our analysis; for example where there were no actors discovered that coordinated across all four events, it may be due to the lack of returns of the Twitter API. 
Additionally, our definition of coordination is limited to the investigation of three single-action coordination, where actors perform the same singular action (e.g. two tweets with the same hashtag) within a short time window. 

\paragraph{Future Work.}
There are several directions for expansions from this study. First, investigation of higher-order actions \cite{DBLP:journals/corr/abs-2105-07454} which are actions combining multiple singular actions, e.g. two tweets with the same hashtag and URL, should be made. Higher order actions combining more than one singular action can provide a source of deliberate coordination rather than coincidental synchronicity. 
Additionally, coordination effects from the three studied aspects -- semantic, referral and social -- coordination can be combined together to build a coordination index to provide a better signal for coordination. One way is to consider these coordination networks as a multiplex network \cite{magnani2021community} before applying community detection methods to sieve out the core structures. 
Nonetheless, we hope this study sheds light on the presence of quantitative evidence of coordinated posting on social media and the user types that participate in these coordinated actions.




\section{Conclusion}
Coordinated groups online have the ability to do real harm to the offline world, as witnessed in the case of the 6 January 2021 Capitol Riots. During the event, far-right groups maintained a strong online presence on multiple social media platforms, spreading propaganda and publicizing the protest, resulting in the violence of riot \cite{atlantic_council_2021}.
In this work, we build on existing approaches of coordination detection using synchronized actions within a time window to further define the choice of an optimal window size for measuring local coordination without capturing coincidental synchronous noise. 
We then perform a comparative study across four case studies demonstrating the presence of coordinated groups on Twitter in four major events in the United States. 
We provide a systematic approach for the discovery and analysis of coordinated groups, defined as groups of users that perform high levels of synchronized actions within a 5-minute sliding window timeframe.
We discover and analyze groups of users that perform semantic, referral and social coordination. 
Through narrative, URL and screen name analysis, we identify themes within network clusters. 
We analyze the users in the groups in terms of their meta-data and quantify the number of ways and number of events the discovered users coordinate. 
We hope that this work will shed light on new techniques that may be used to identify coordinated groups of users on social media, and will be adapted to sieve out malicious coordinating groups attempting to influence the network.

\newpage
\section*{Acknowledgments}
We would like to thank Thomas Magelinski for his inputs to this paper. 
The research for this paper was supported in part by the Knight Foundation and the Office of Naval Research Grant (N000141812106) by the center for Informed Democracy and Social-cybersecurity  (IDeaS) and the center for Computational Analysis of Social and Organizational Systems (CASOS) at Carnegie Mellon University. The views and conclusions  are those of the authors and should not be interpreted as representing the official  policies, either expressed or implied, of the Knight Foundation, Office of Naval Research or the US Government.

\bibliographystyle{ACM-Reference-Format}
\bibliography{references}

\end{document}